\documentclass[twocolumn]{aastex61}

\received{July 17, 2017}
\revised{? ?, 2017}
\accepted{November 9, 2017}
\submitjournal{ApJ}

\shorttitle{An evolving synchrotron jet spectrum in Swift J1357.2--0933}
\shortauthors{Russell et al.}

\begin{document}

\title{Optical Precursors to Black Hole X-ray Binary Outbursts: An evolving synchrotron jet spectrum in Swift J1357.2--0933}

\author{David M. Russell}
\affiliation{New York University Abu Dhabi, P.O. Box 129188, Abu Dhabi, UAE}
\email{dave.russell@nyu.edu}

\author{Ahlam Al Qasim}
\affiliation{New York University Abu Dhabi, PO Box 129188, Abu Dhabi, UAE}

\author{Federico Bernardini}
\affiliation{New York University Abu Dhabi, PO Box 129188, Abu Dhabi, UAE}
\affiliation{INAF -- Osservatorio Astronomico di Roma, via Frascati 33, I-00040 Monteporzio Catone (Roma), Italy}

\author{Richard M. Plotkin}
\affiliation{International Centre for Radio Astronomy Research -- Curtin University, GPO Box U1987, Perth, WA 6845, Australia}

\author{Fraser Lewis}
\affiliation{Faulkes Telescope Project, School of Physics, and Astronomy, Cardiff University, The Parade, Cardiff, CF24 3AA, Wales, UK}
\affiliation{Astrophysics Research Institute, Liverpool John Moores University, 146 Brownlow Hill, Liverpool L3 5RF, UK}

\author{Karri I. I. Koljonen}
\affiliation{Finnish Centre for Astronomy with ESO (FINCA), University of Turku, V\"ais\"al\"antie 20, FI-21500 Piikki\"o, Finland}
\affiliation{Aalto University Mets\"ahovi Radio Observatory, P.O. Box 13000, FI-00076 Aalto, Finland}

\author{Yi-Jung Yang}
\affiliation{School of Physics and Astronomy, Sun Yat-Sen University, 135 Xingang Xi Road, Guangzhou 510275, People's Republic of China}



\begin{abstract}
We present six years of optical monitoring of the black hole (BH) candidate X-ray binary Swift J1357.2--0933, during and since its discovery outburst in 2011. On these long timescales, the quiescent light curve is dominated by high amplitude, short-term (seconds--days) variability spanning $\sim 2$ mag, with an increasing trend of the mean flux from 2012 to 2017 that is steeper than in any other X-ray binary found to date (0.17 mag yr$^{-1}$). We detected the initial optical rise of the 2017 outburst of Swift J1357.2--0933, and we report that the outburst began between 2017 April 1 and 6. Such a steep optical flux rise preceding an outburst is expected according to disk instability models, but the high amplitude variability in quiescence is not. Previous studies have shown that the quiescent spectral, polarimetric, and rapid variability properties of Swift J1357.2--0933 are consistent with synchrotron emission from a weak compact jet. We find that a variable optical/infrared spectrum is responsible for the brightening: a steep, red spectrum before and soon after the 2011 outburst evolves to a brighter, flatter spectrum since 2013. The evolving spectrum appears to be due to the jet spectral break shifting from the infrared in 2012 to the optical in 2013, then back to the infrared by 2016--2017 while the optical remains relatively bright. Swift J1357.2--0933 is a valuable source to study BH jet physics at very low accretion rates and is possibly the only quiescent source in which the optical jet properties can be regularly monitored.
\end{abstract}




\section{Introduction} \label{sec:intro}

It has been known for more than a decade that accreting black holes (BHs) can launch jets at very low accretion rates, when the X-ray luminosities are less than $\sim 10^{-5}$ of the Eddington luminosity. Radio emission has been detected from jets released by low-mass X-ray binaries (LMXBs) in quiescence (with X-ray luminosities $\sim 10^{30}$--$10^{33.5}$ erg s$^{-1}$), in a growing number of systems, all hosting BH candidates \citep[e.g.][]{hjelet00,gallet05,gallet06,gallet14,millet11,dzibet15,market15}. LMXBs spend most of their time in quiescence between outbursts, and therefore presumably (at least in the BH systems) launch jets for most of their lifetimes. However, long-term radio studies of LMXB jets in quiescence do not exist to date, because they are generally too faint to monitor with current radio facilities. Most BH systems possess $\mu$Jy-level flux densities, with typically only a few detections of a source made over decades \citep[e.g.][]{gallet06,gallet14,riboet17}. Consequently, the long-term behavior and properties of quiescent jets from Galactic BHs remain poorly known. 

At optical wavelengths, long-term monitoring studies of quiescent LMXBs have revealed the ellipsoidal modulation of the companion star in some systems, leading to measurements of the fundamental system parameters such as the masses and orbital inclinations \citep*[see][for a review]{casajo14}. In other LMXBs, optical flickering, flares and/or variability are seen from the accretion flow \citep[e.g.][]{yanget12,koljet16,wuet16} and some exhibit a combination of the above contributions \citep[e.g.][]{zuriet03,shahet05,cantet10,macdet14,bernet16}. Theoretically, the disk instability model \citep*[DIM; e.g.][]{dubuet01,laso01,hameet17} predicts that between outbursts, the temperature and surface density of the accretion disk increase as matter builds up in the disk, leading to higher optical fluxes. This phenomenon has only recently been seen in LMXBs with long-term (years) optical monitoring in quiescence \citep{yanget12,bernet16,koljet16,wuet16}.

In one BH system, Swift J1357.2--0933, a deep radio observation yielded a $3\sigma_{\rm rms}$ upper limit of 3.9 $\mu$Jy during quiescence, whereas synchrotron emission, likely from the jet, was detected at optical--infrared (OIR) wavelengths \citep{plotet16}. While OIR synchrotron emission has been detected in a number of BH LMXBs during quiescence \citep*[e.g.][]{gallet07,geliet10,russet13} and one neutron star system \citep{baglet13}, only in Swift J1357.2--0933 does it appear to dominate the quiescent OIR spectrum. High amplitude seconds to hours-timescale optical variability, a red or flat spectral energy distribution (SED), and evidence for intrinsic polarization \citep{shahet13,plotet16,russet16} are all unique properties of Swift J1357.2--0933. These properties cannot be produced by the underlying accretion flow or the companion star. However, emission from the jet can account for all of these properties. This has led to the conclusion that jets are continuously launched during quiescence.
\cite{plotet16} indeed found that the OIR SED was flat ($\alpha \approx 0$, where $F_{\nu} \propto \nu^{\alpha}$), and turned over to a steeper slope at shorter wavelengths in the optical--UV regime, consistent with the spectral break in the jet spectrum between optically thin and optically thick, partially self-absorbed synchrotron emission. This break has been identified in a number of LMXBs during outbursts (see \citealt{russet13} and references therein, and more recently \citealt{russet14,koljet15,diazet17}). Swift J1357.2--0933 has provided the only robust measurement of a `jet break' for a quiescent X-ray binary.

Here, we present six years of optical monitoring of Swift J1357.2--0933, leading up to the recently discovered 2017 outburst of the source \citep{dincet17,draket17,sivaet17}.
We detect the outburst rise at an earlier stage than previously reported by \cite{draket17}. 
We also collected all OIR data available to investigate the long-term quiescent behavior of the OIR flux and SED of Swift J1357.2--0933, before and since its discovery outburst in 2011 \citep{krimet11a}.

\begin{figure*}
\centering
\includegraphics[height=11cm,angle=0]{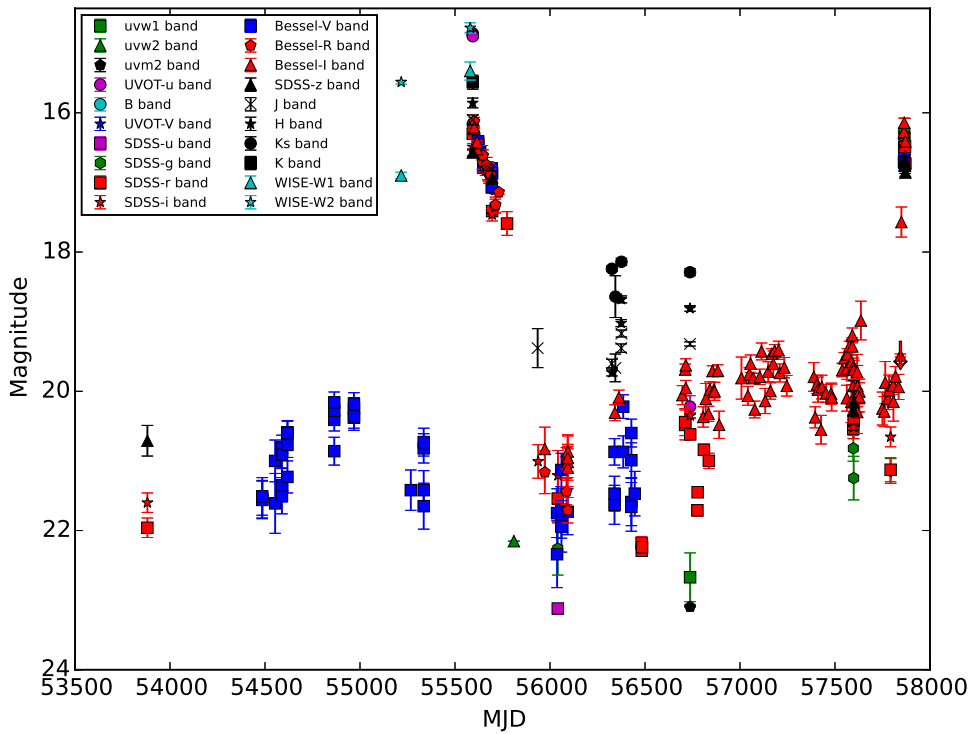}
\caption{OIR light curves of Swift J1357.2--0933 from 2006 to 2017. The 2011 and 2017 outbursts are evident near the center and right, respectively. Large amplitude variability is present in quiescence before and between the two outbursts.}
\end{figure*}

\section{Data collection}

\subsection{Faulkes Telescope monitoring}

We have conducted a long-term monitoring campaign of Swift J1357.2--0933 with the two, robotic 2~m Faulkes Telescopes (North, at Haleakala on Maui, Hawaii, USA, and South, at Siding Spring, Australia) since its 2011 outburst. Our observations are part of an ongoing monitoring campaign of $\sim 40$ LMXBs \citep{lewiet08}. Most observations were made using the Bessell $I$-band filter, with some in Bessell $R$-band in 2011--2012, and some Sloan Digital Sky Survey (SDSS) $g^{\prime}$, $r^{\prime}$, $i^{\prime}$, $z^{\prime}$ consecutive observations in 2016--2017. The latter were made specifically to investigate the optical SED and measure the spectral index. 
Both Faulkes Telescopes are equipped with cameras with a pixel scale of 0.30~arcsec pixel$^{-1}$ and a field of view of $10 \times 10$ arcmin, except in 2011 February, in which the cameras had 0.28 arcsec pixel$^{-1}$ and 4.8 arcmin field of view. 
We detected the source in a total of 103 images between 2011 February and 2017 March, 79 of which were taken in the Bessell $I$-band filter.

We also present monitoring of the rise of the 2017 outburst. Data were taken in 2017 April with the 2~m Faulkes Telescopes as well as some of the 1~m network Las Cumbres Observatory (LCO) telescopes: those at Cerro Tololo (Chile) and the South African Astronomical Observatory (SAAO; Sutherland, South Africa). The filters $u^{\prime}$, $g^{\prime}$, $r^{\prime}$, $i^{\prime}$, $z^{\prime}$, and $I$ were used, and the 1~m telescopes were equipped with cameras with 0.39~arcsec pixel$^{-1}$ and a field of view of $\sim 26.5 \times 26.5$ arcmin.

Photometry was carried out using \textsc{PHOT} in \textsc{IRAF}.\footnote{IRAF is distributed by the National Optical Astronomy Observatory, which is operated by the Association of Universities for Research in Astronomy, Inc., under cooperative agreement with the National Science Foundation.} Flux calibration was achieved using the SDSS magnitudes of several stars in the field, from SDSS Data Release 12 \citep{alamet15}. We derived the Bessell $I$-band and $R$-band magnitudes of the field stars using their SDSS magnitudes, adopting the color transformations of \cite*{jordet06}. We reported one of our $I$-band magnitudes in \cite{russet16} that was quasi-simultaneous with near-infrared (near-IR) polarimetric observations; all other Faulkes data are new to this paper.

\subsection{Archival data from transient surveys}

Swift J1357.2--0933 has 52 optical $V$-band detections in the Catalina Real-Time Transient Survey \citep[CRTS-I;][]{draket09} from 2008 January until 2013 June (including some detections of the outburst in 2011). The CRTS Transient ID of the source is MLS110301:135717-093239 \citep[see also][]{draket17}. Swift J1357.2--0933 was also observed by the intermediate Palomar Transient Factory (iPTF) catalog \citep{ofeket12} on 2014 February 23 (during quiescence). Two observations were made using the PTF \emph{Mould-R} filter, which is similar in shape to the SDSS $r^{\prime}$-band filter, but shifted 27~$\AA$ redwards. This is a small difference, and we treat these as SDSS $r^{\prime}$-band magnitudes. We performed aperture photometry on the PTF images using an aperture radius of 3 pixels (3 arcsec). The source is detected in both images.

\subsection{Archival data from the literature}

A search of the literature was performed to gather OIR photometry measurements during quiescence (and some during outburst). The data \citep*[found in][]{rauet11,krimet11b,corret13,shahet13,armset14,mataet15,wengzh15,plotet16,russet16} span a wavelength range from 193 nm in the near-UV to 4.6 $\mu$m in the IR.
Other instruments and telescopes with detections of Swift J1357.2--0933 are the 2.5 m SDSS telescope at Apache Point Observatory in New Mexico (USA), the 2.2 m MPI/ESO telescope at La Silla Observatory (Chile) equipped with the Gamma-Ray Burst Optical/Near-Infrared Detector (GROND) instrument, the Instituto de Astrof\'isica de Canarias (IAC) 0.82 m IAC80 telescope at the Teide Observatory in Tenerife (Spain), and several telescopes located at Roque de Los Muchachos Observatory, La Palma (Spain): the 10.4 m Gran Telescopio Canarias (GTC), the 4.2 m William Herschel Telescope (WHT), the 2.6 m Nordic Optical Telescope (NOT), the 2.5 m Isaac Newton Telescope (INT), the 2.0 m Liverpool Telescope (LT), and the 1.2 m Mercator Telescope (MT). 
We also include the first reported detection of the 2017 outburst, by CRTS-II in $V$-band \citep{draket17}. All magnitudes from the literature, transient surveys, and our Faulkes Telescope monitoring are shown in Fig. 1.

\section{Results and analysis}

Most LMXBs vary in quiescence by a few tenths of a magnitude, due to the ellipsoidal modulation of the companion star, and/or weak accretion activity. In Fig. 1, we see that Swift J1357.2--0933 exhibits long-term, very high amplitude OIR variations during quiescence. The $I$-band magnitudes span a range $I \sim 21.1$ to $I \sim 19.0$ during quiescence (changes of a factor of seven in flux), which is a much greater amplitude than expected from ellipsoidal modulations of the companion star. The majority of the emission must therefore be produced in the accretion flow, jet, or by X-ray reprocessing. While our Faulkes Telescope monitoring began during the 2011 outburst, many of the CRTS detections preceded it, and $V$-band variations spanning almost 2 mag are present at these earlier times. We still consider the source to be in a state of quiescence despite the strong variations, because the X-ray luminosity is very low; $L_{\rm X}$ (0.5-10 keV) $\sim 8 \times 10^{29}$ -- $1 \times 10^{31}$ erg s$^{-1}$ in 2013 July, depending on the distance to the source \citep{armset14}.

\begin{figure}
\centering
\includegraphics[height=7.05cm,angle=0]{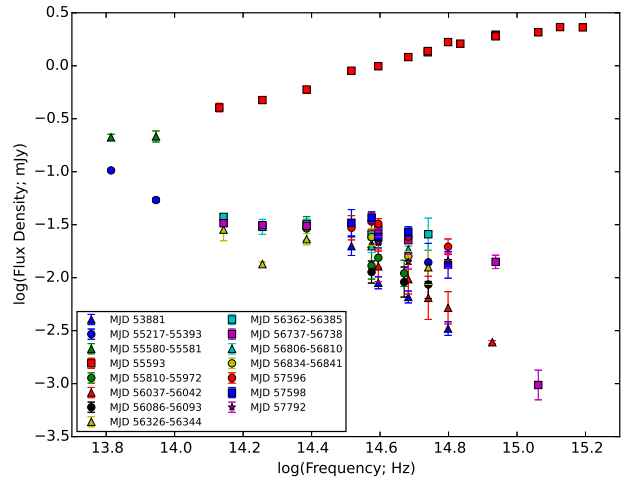}
\caption{OIR SEDs constructed from various date ranges as shown in the key. All SEDs are from periods of quiescence except MJD 55580--55581 and MJD 55593, which were during the early stages of the 2011 outburst.}
\end{figure}

\begin{figure*}
\centering
\includegraphics[height=4.2cm,angle=0]{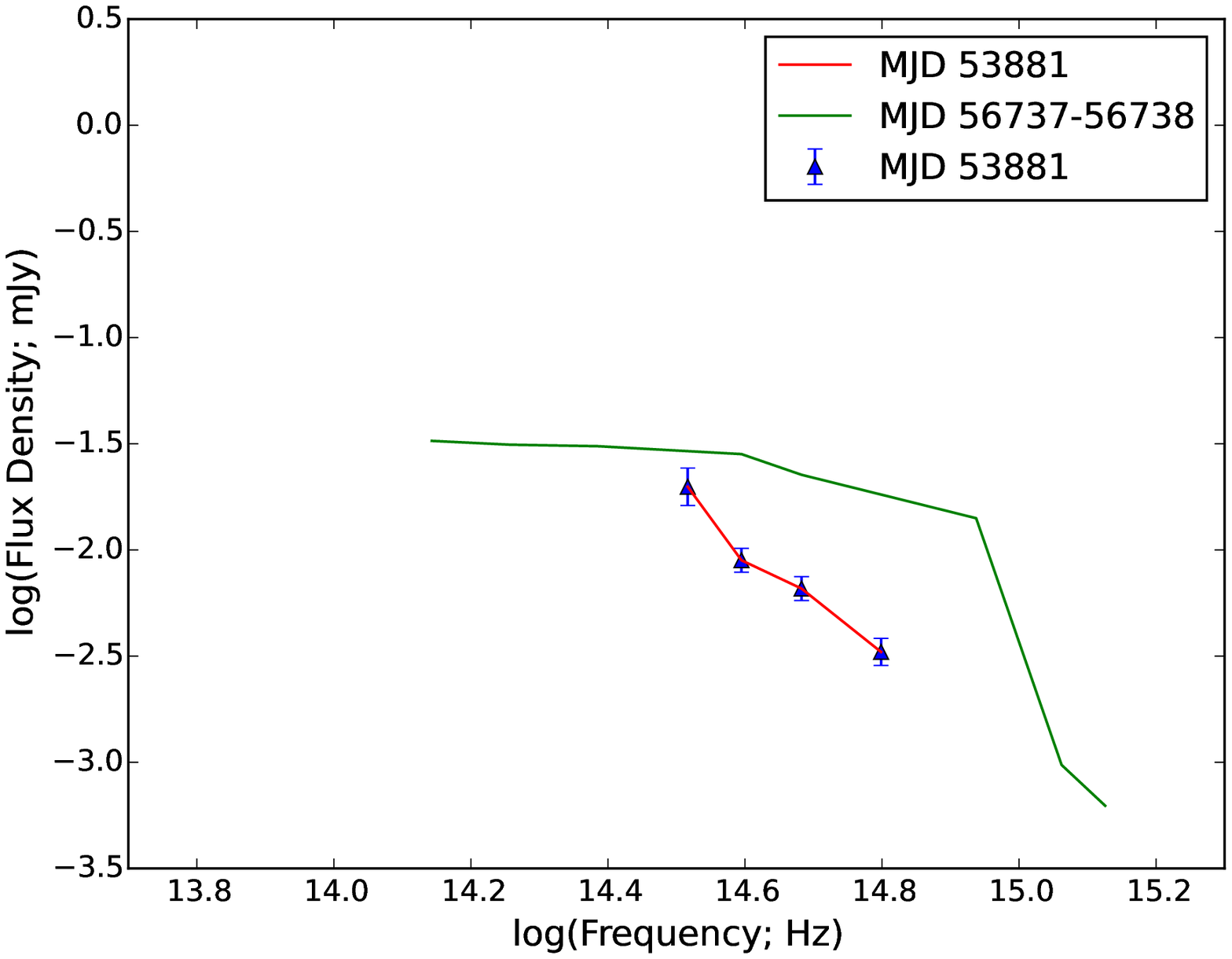}
\includegraphics[height=4.2cm,angle=0]{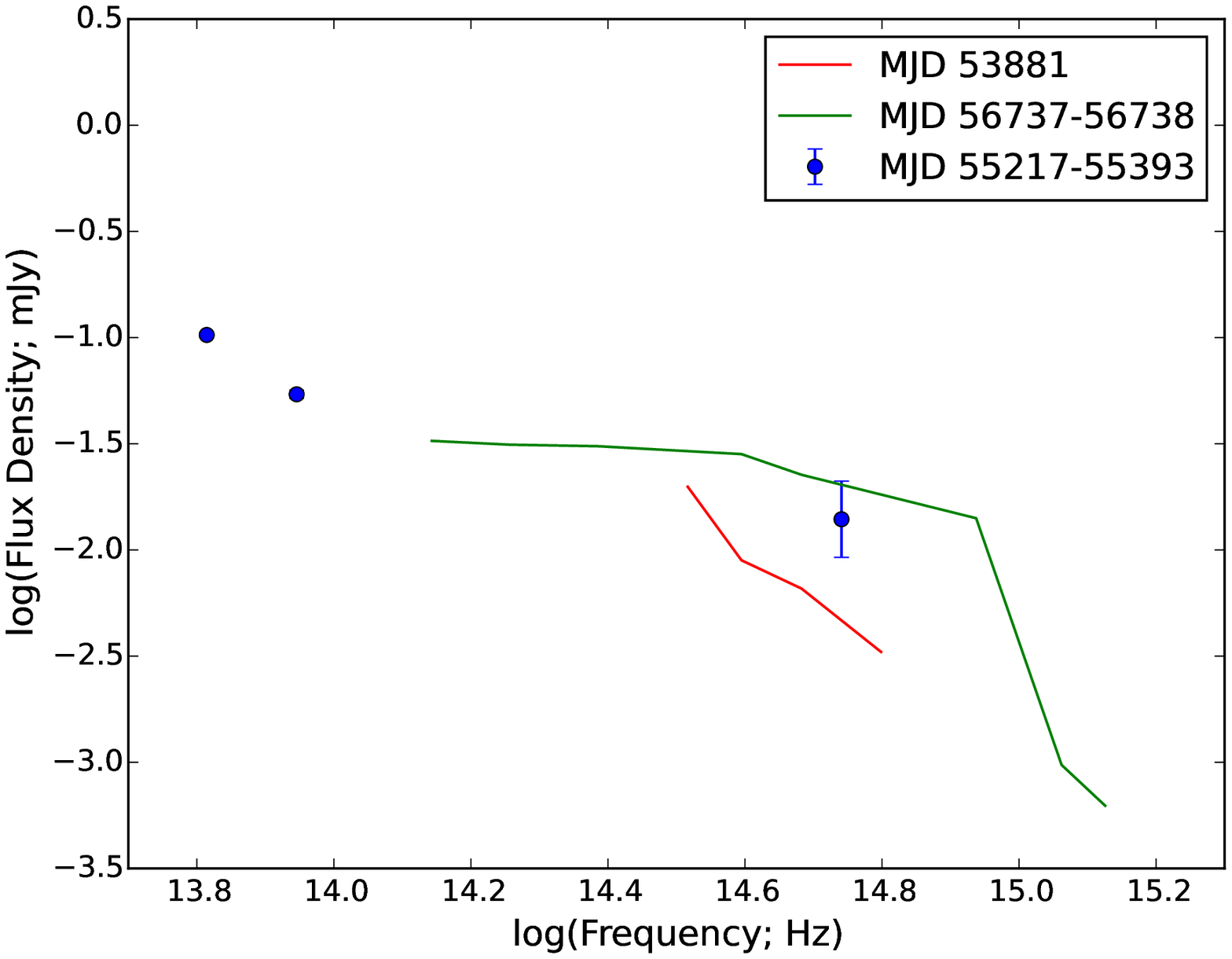}
\includegraphics[height=4.2cm,angle=0]{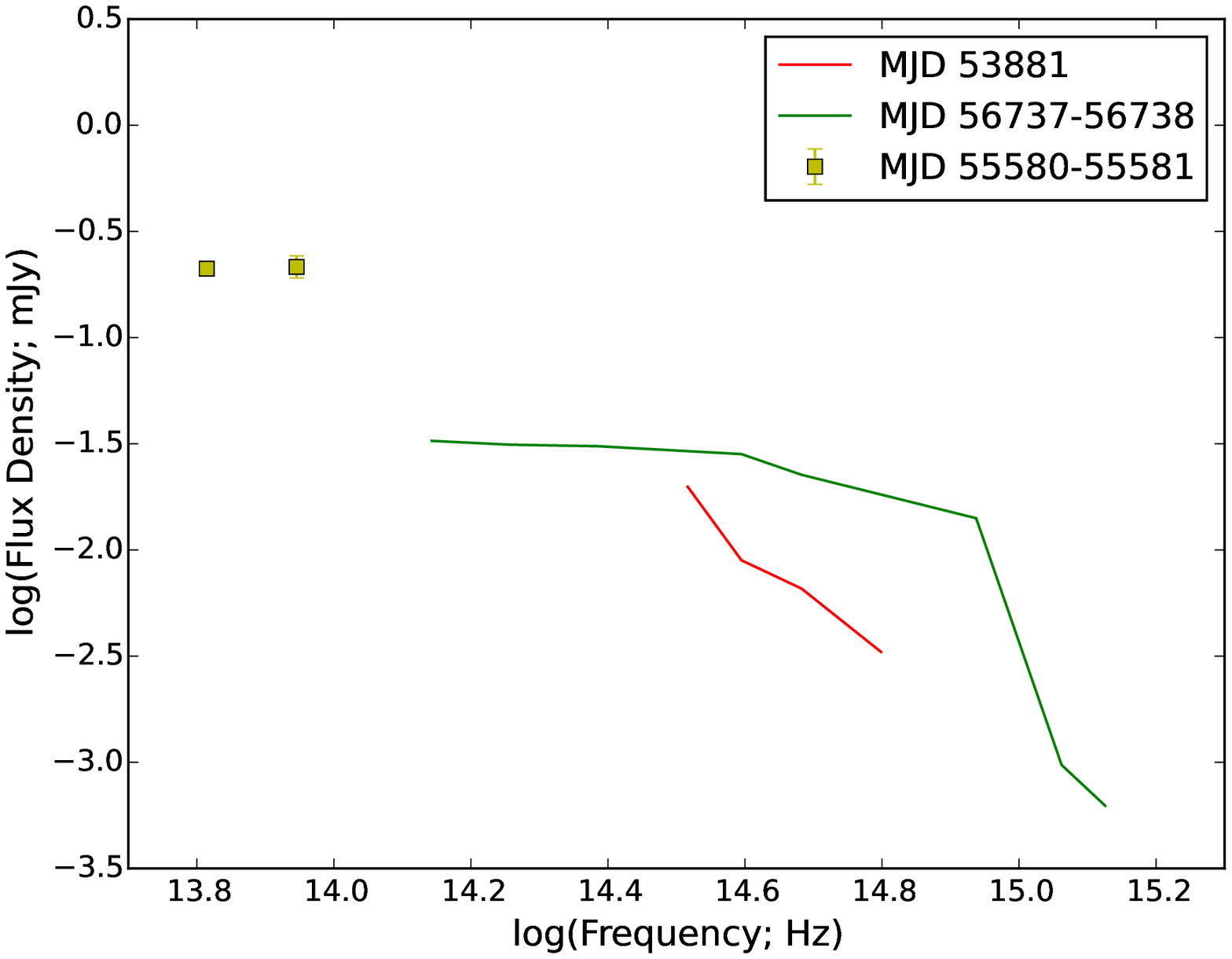} \\
\includegraphics[height=4.2cm,angle=0]{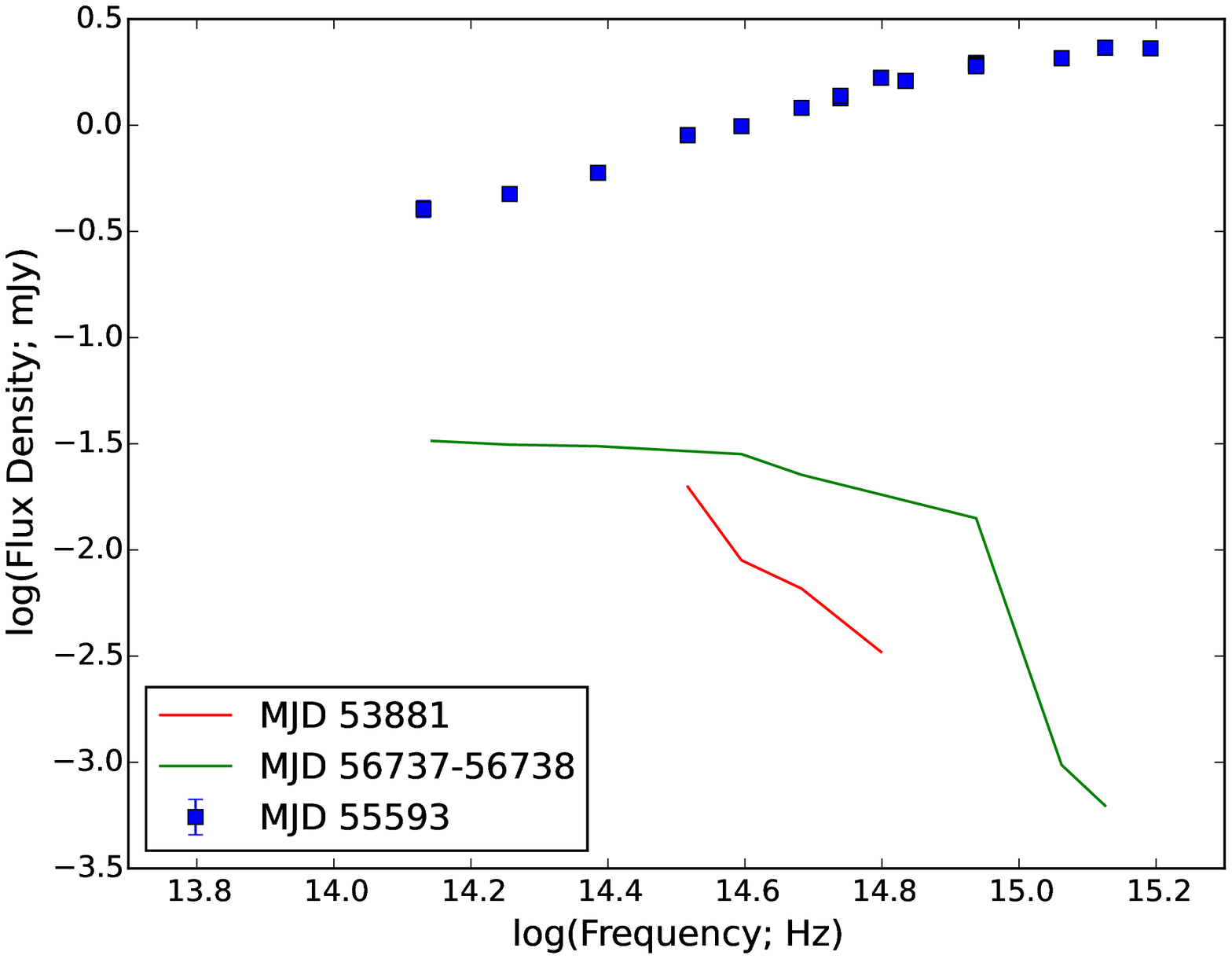}
\includegraphics[height=4.2cm,angle=0]{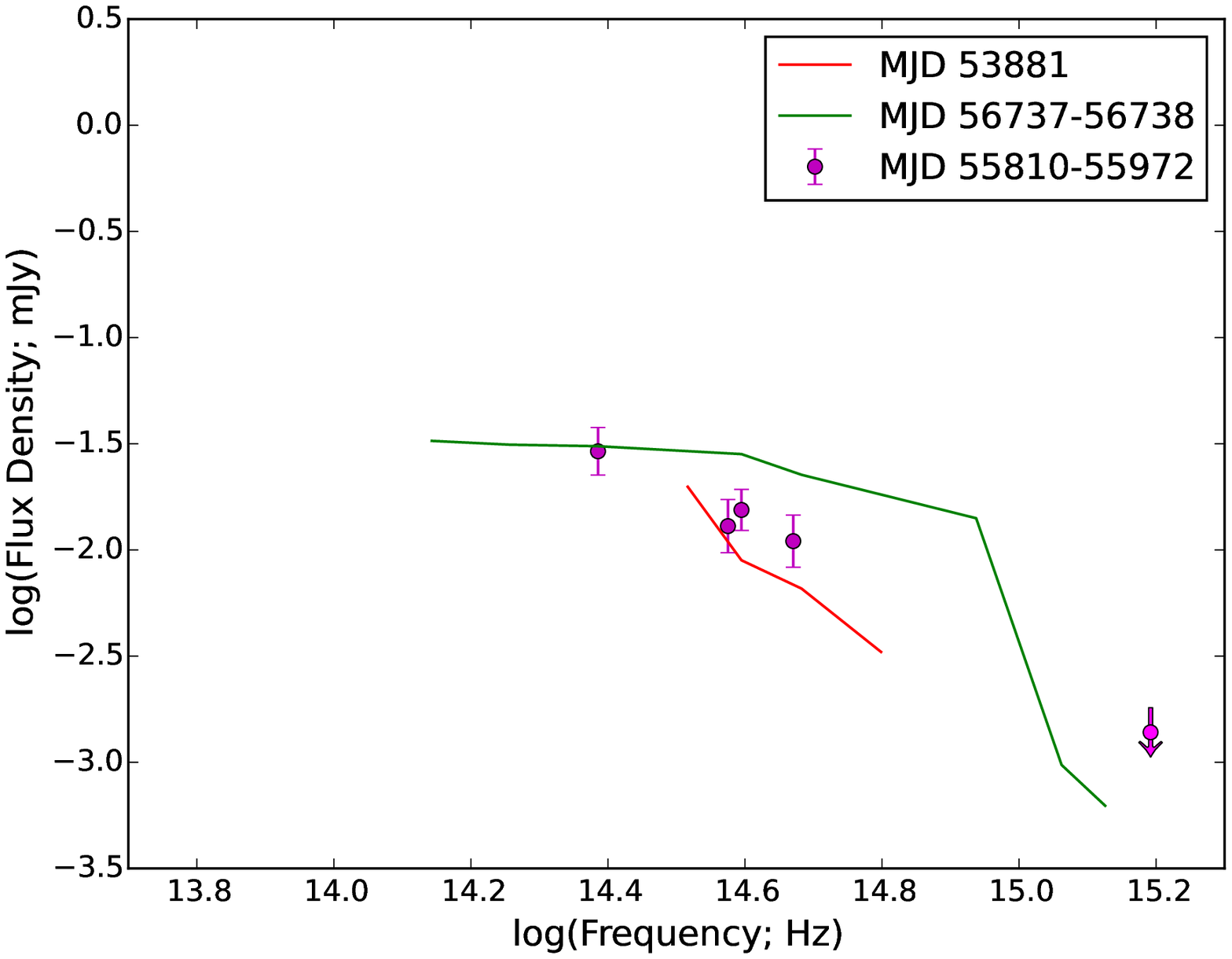}
\includegraphics[height=4.2cm,angle=0]{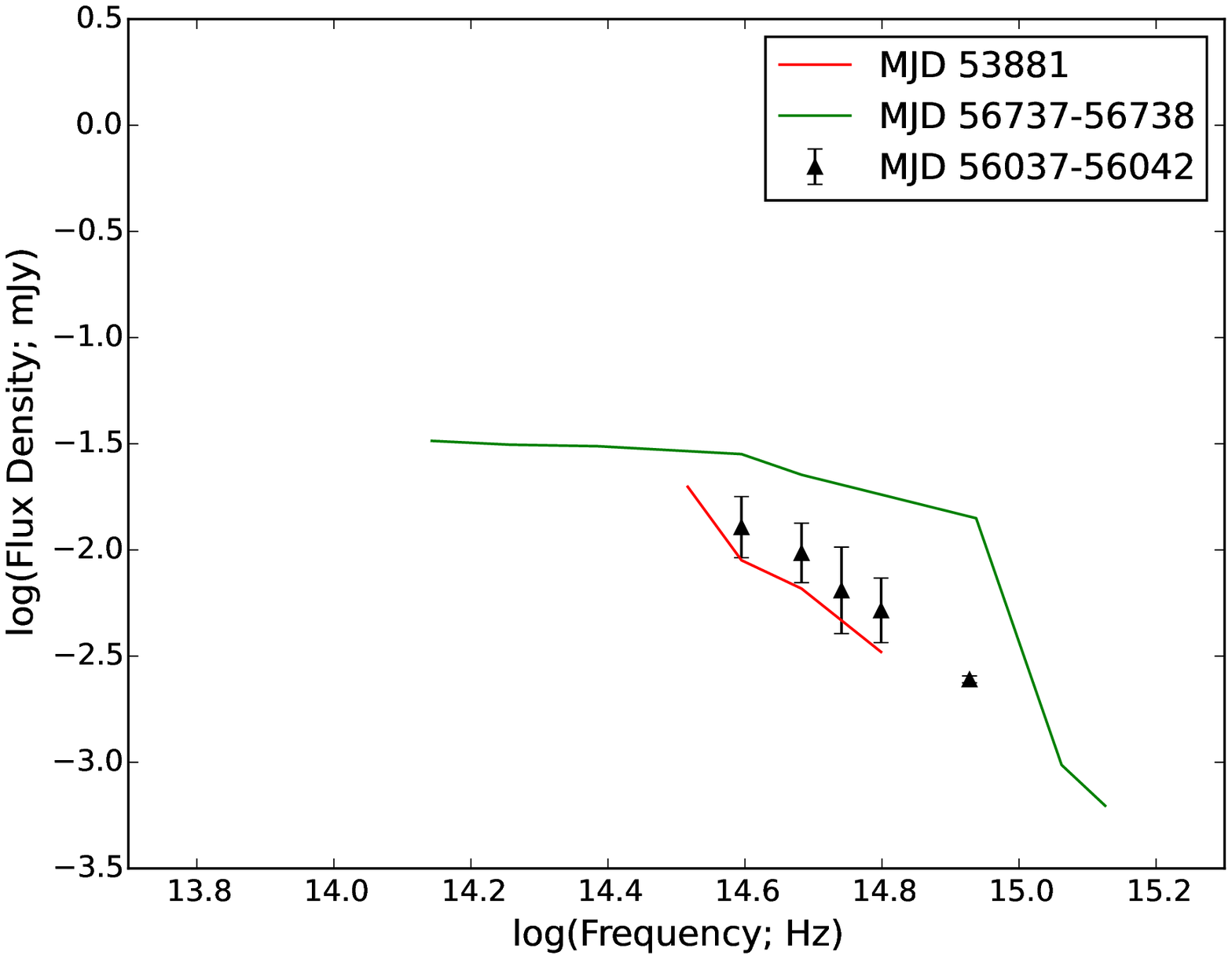} \\
\includegraphics[height=4.2cm,angle=0]{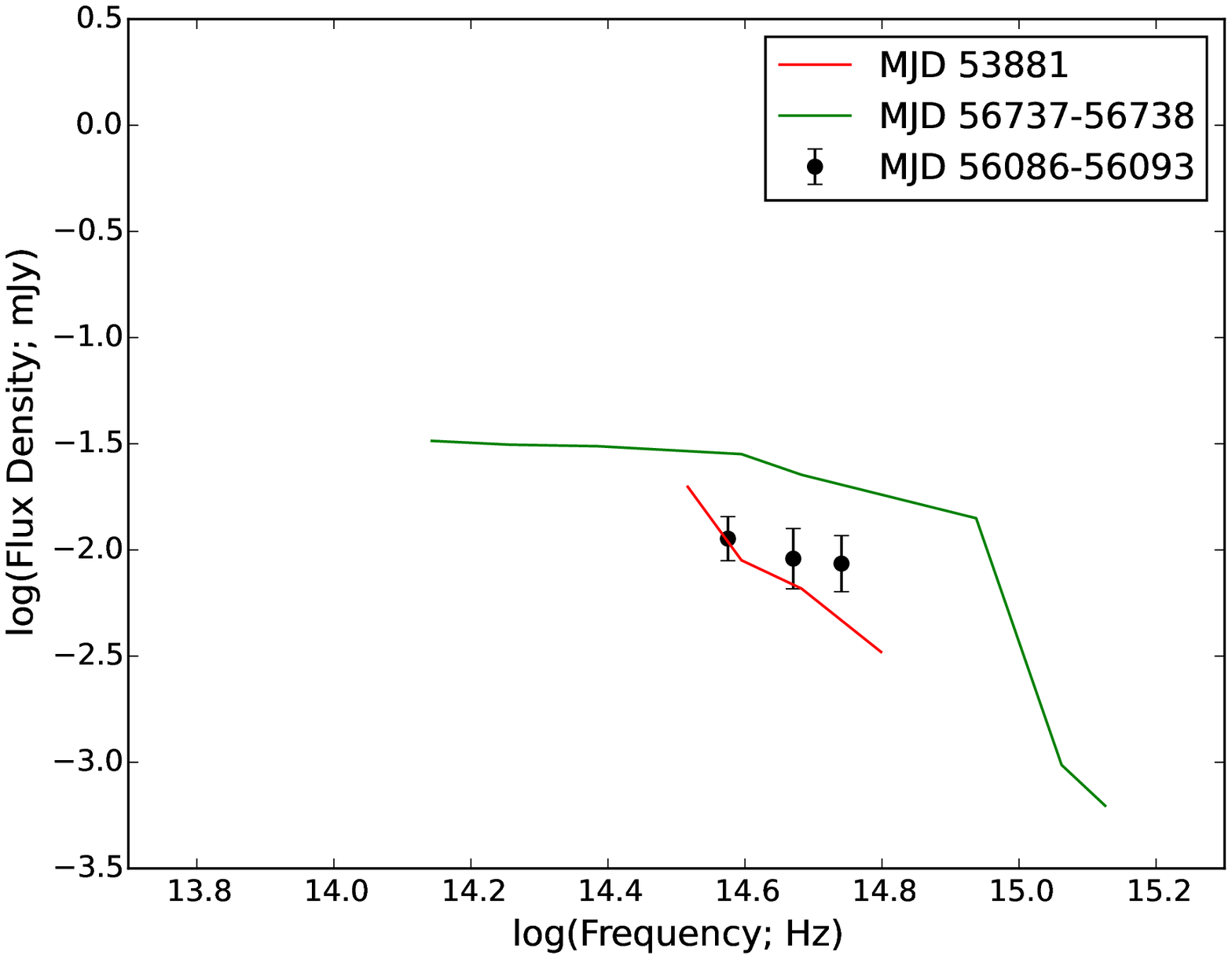}
\includegraphics[height=4.2cm,angle=0]{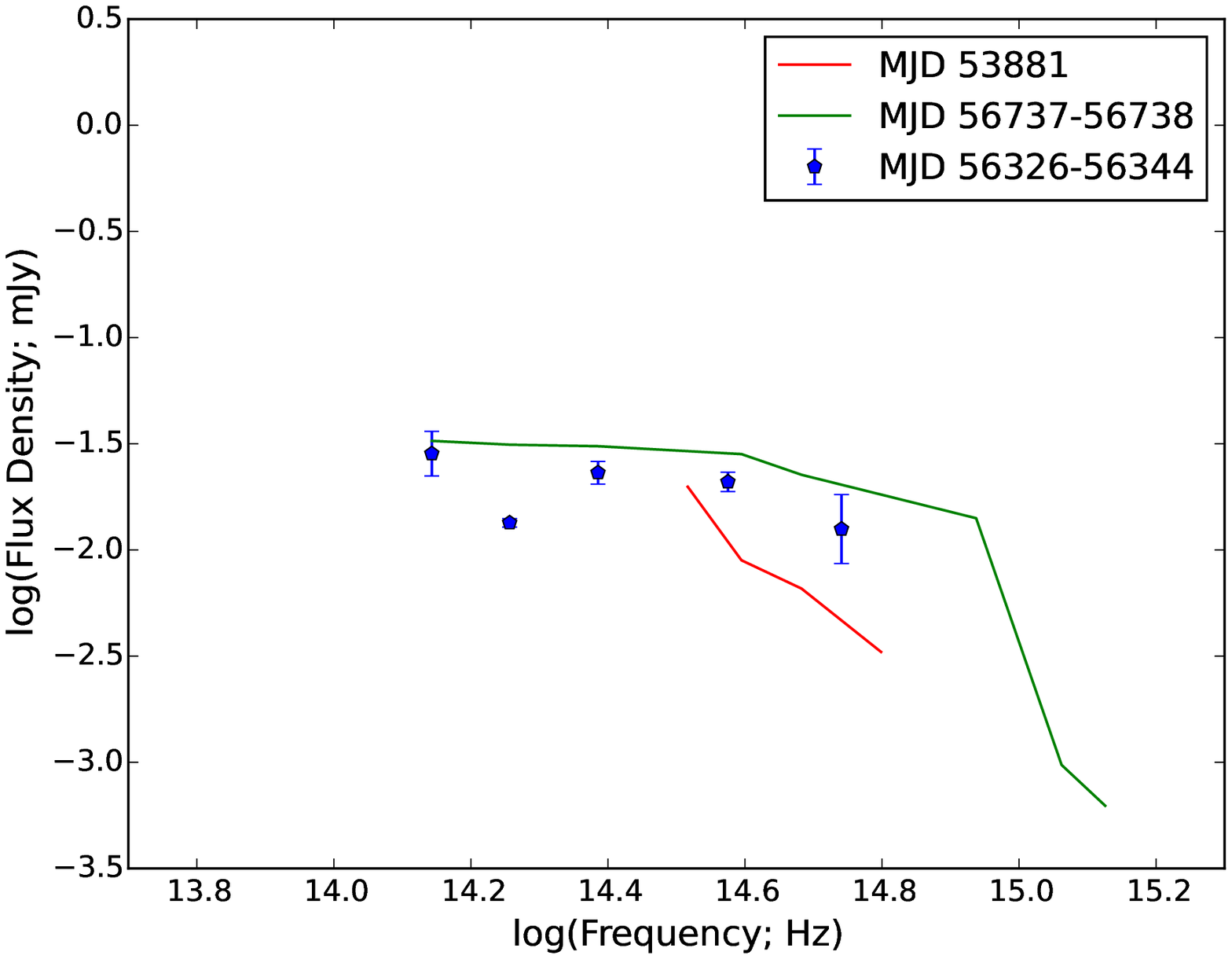}
\includegraphics[height=4.2cm,angle=0]{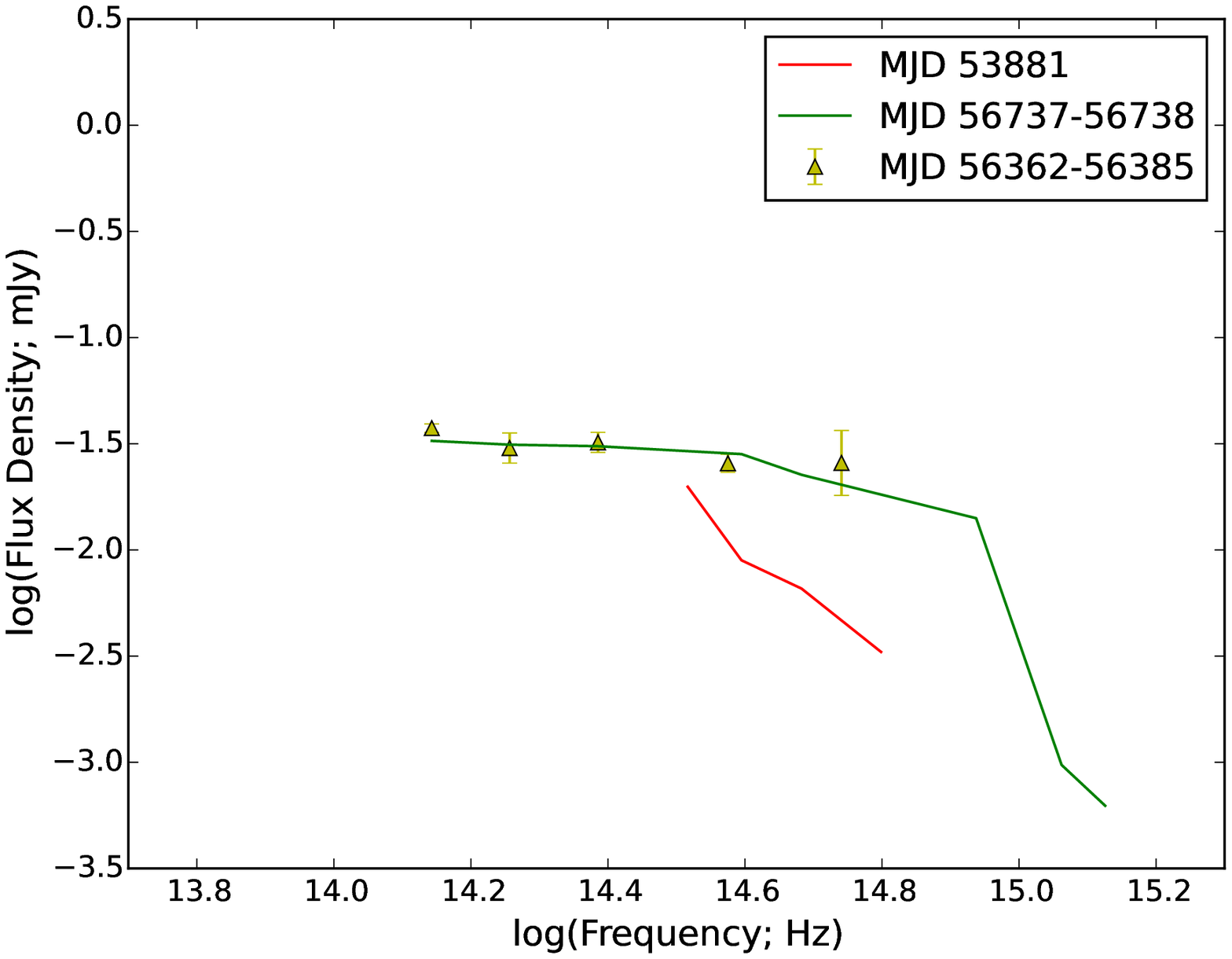} \\
\includegraphics[height=4.2cm,angle=0]{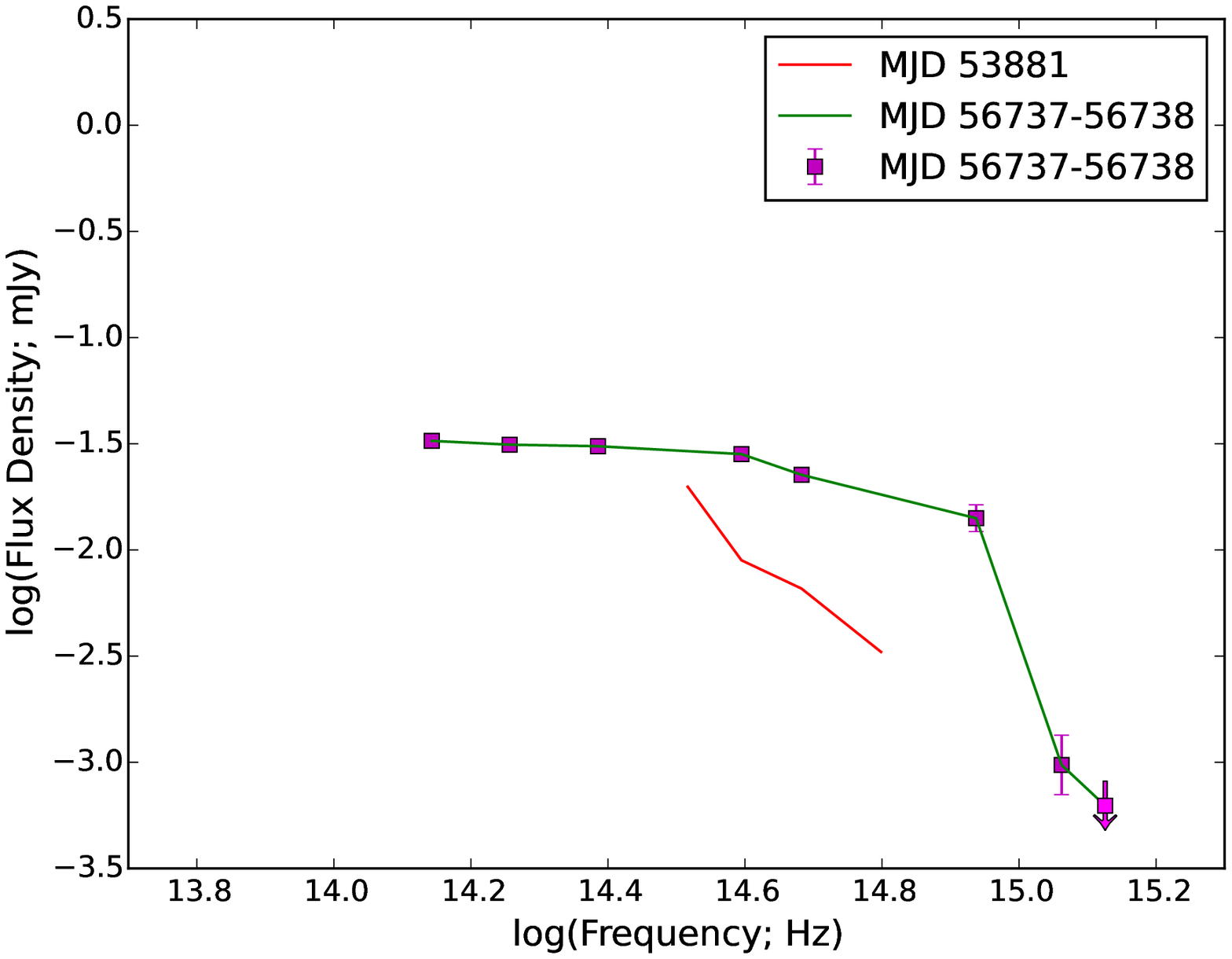}
\includegraphics[height=4.2cm,angle=0]{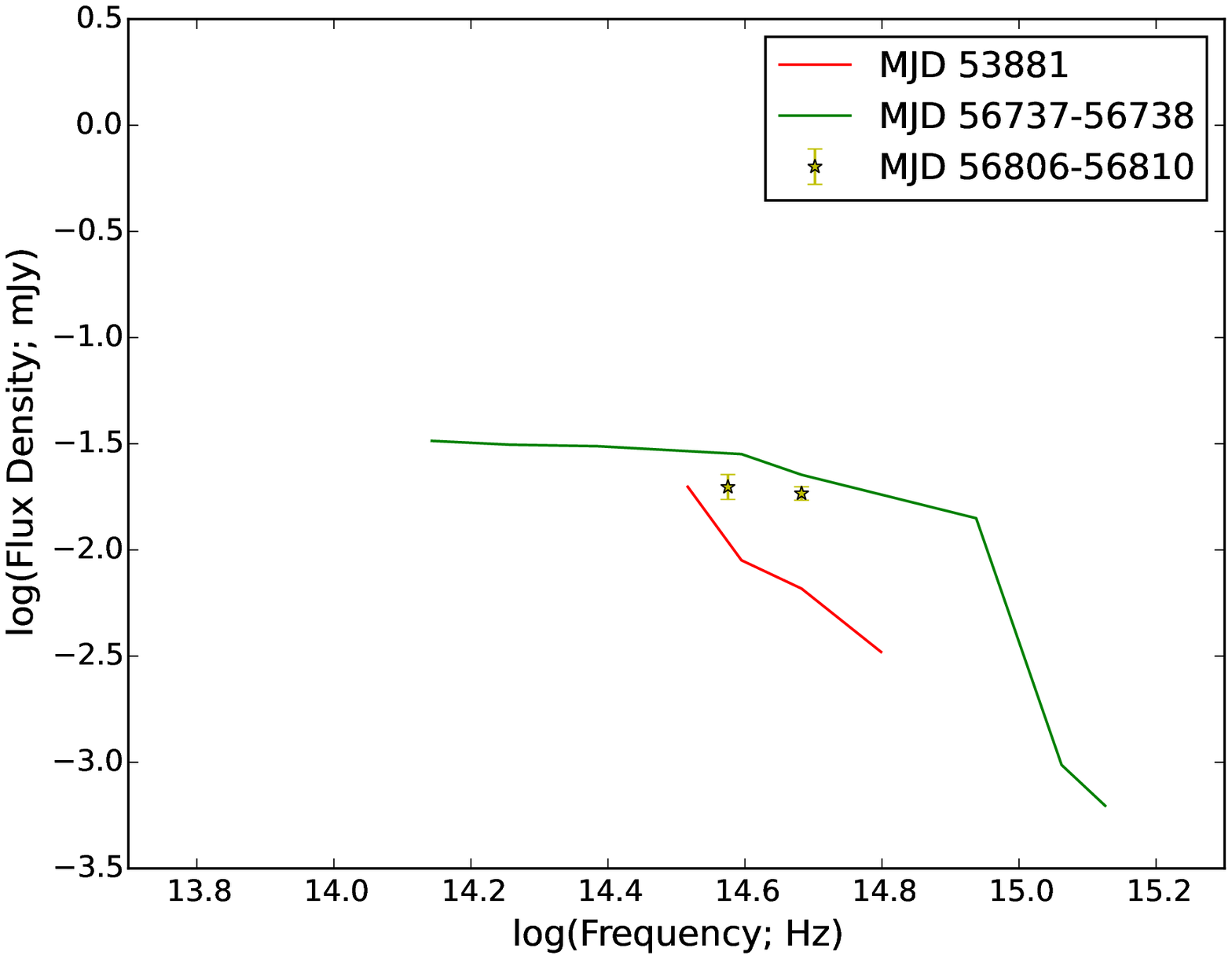}
\includegraphics[height=4.2cm,angle=0]{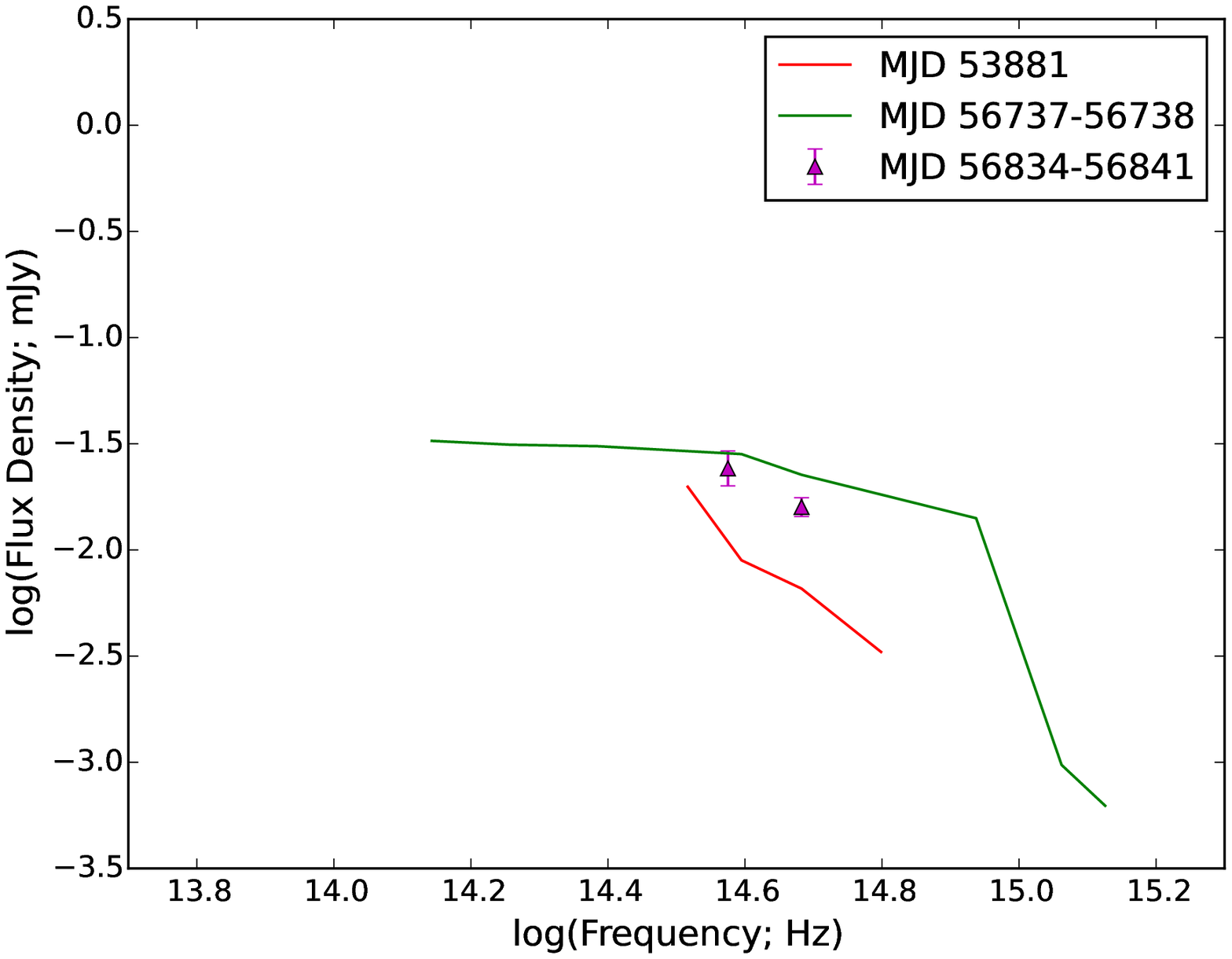} \\
\includegraphics[height=4.2cm,angle=0]{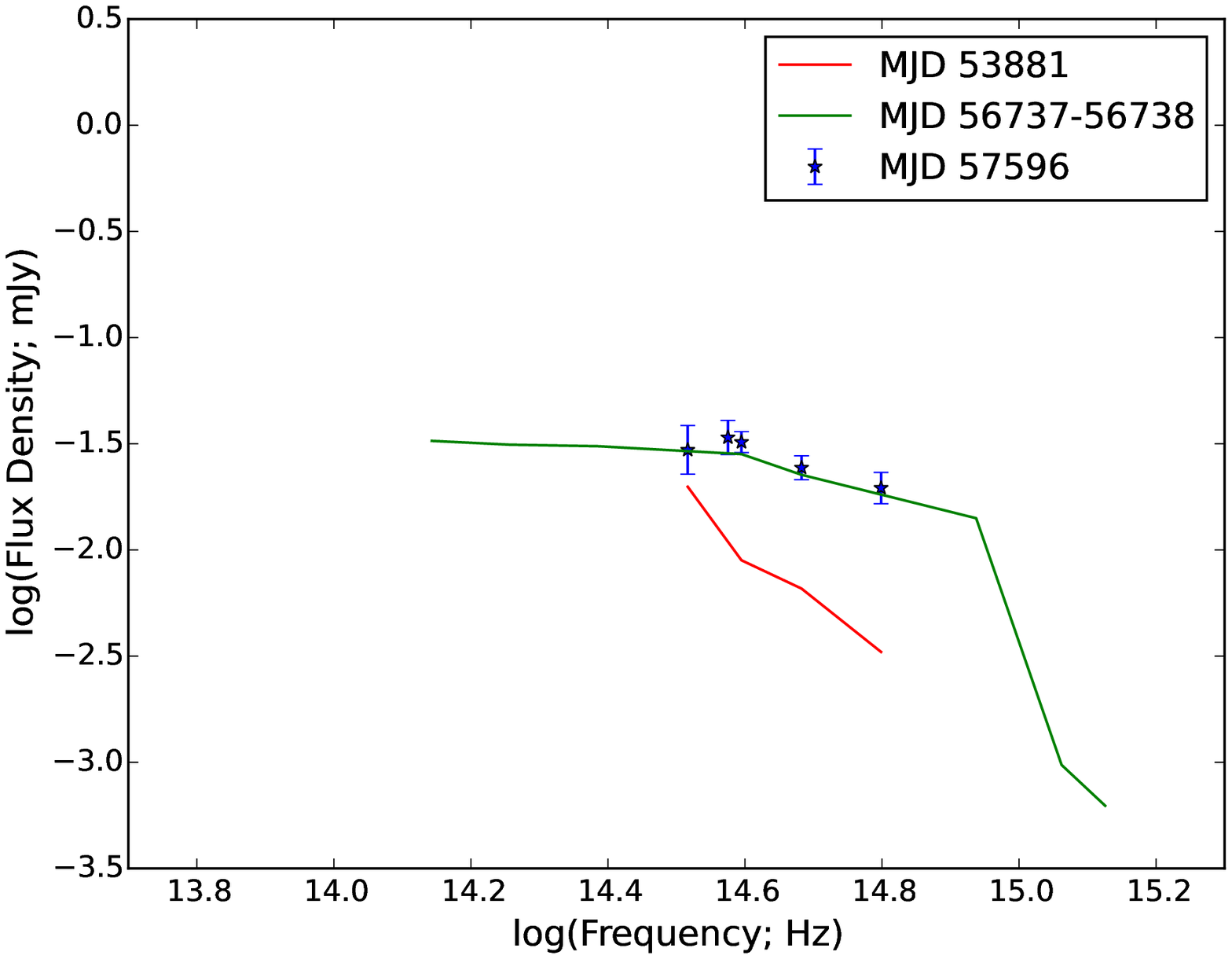}
\includegraphics[height=4.2cm,angle=0]{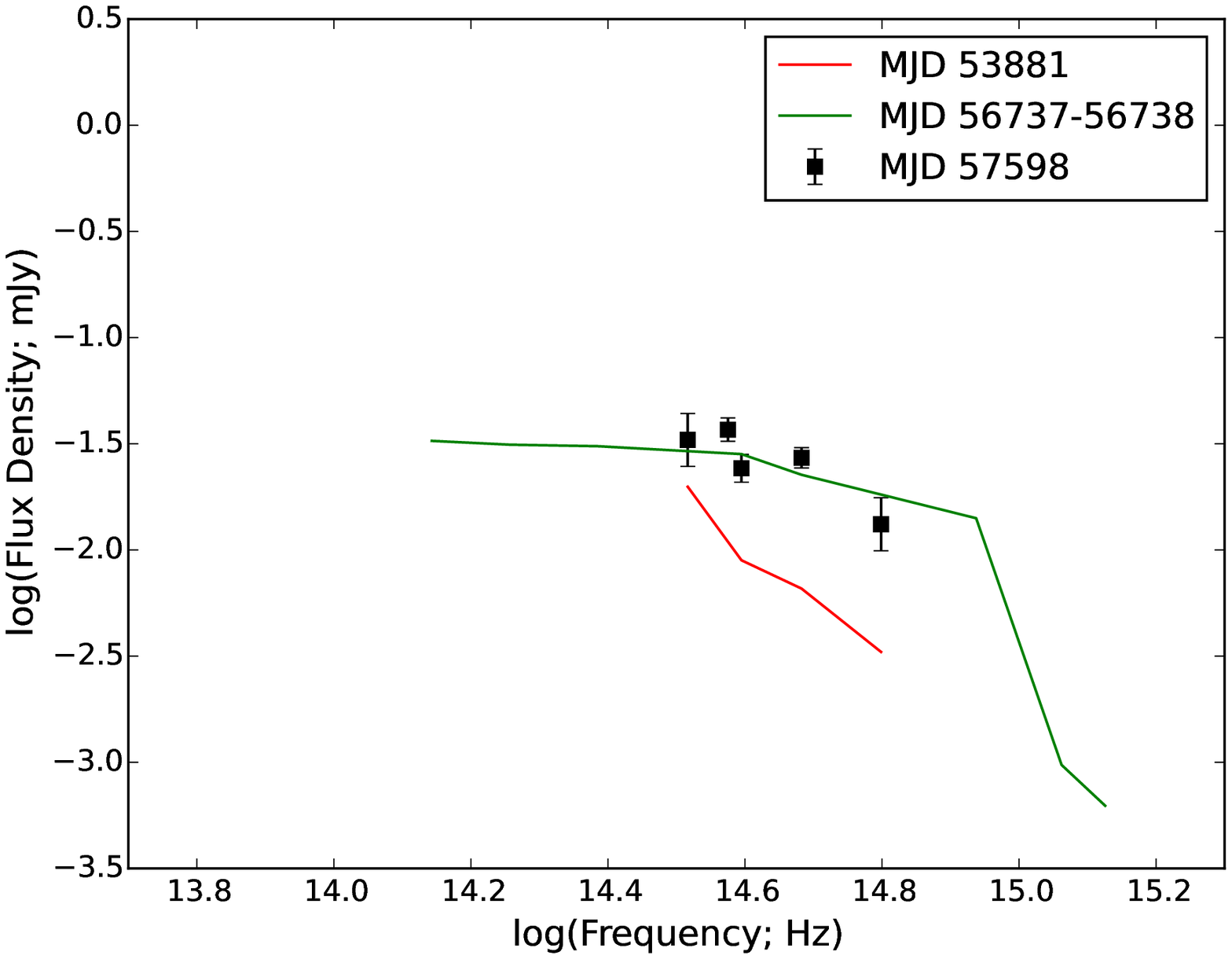}
\includegraphics[height=4.2cm,angle=0]{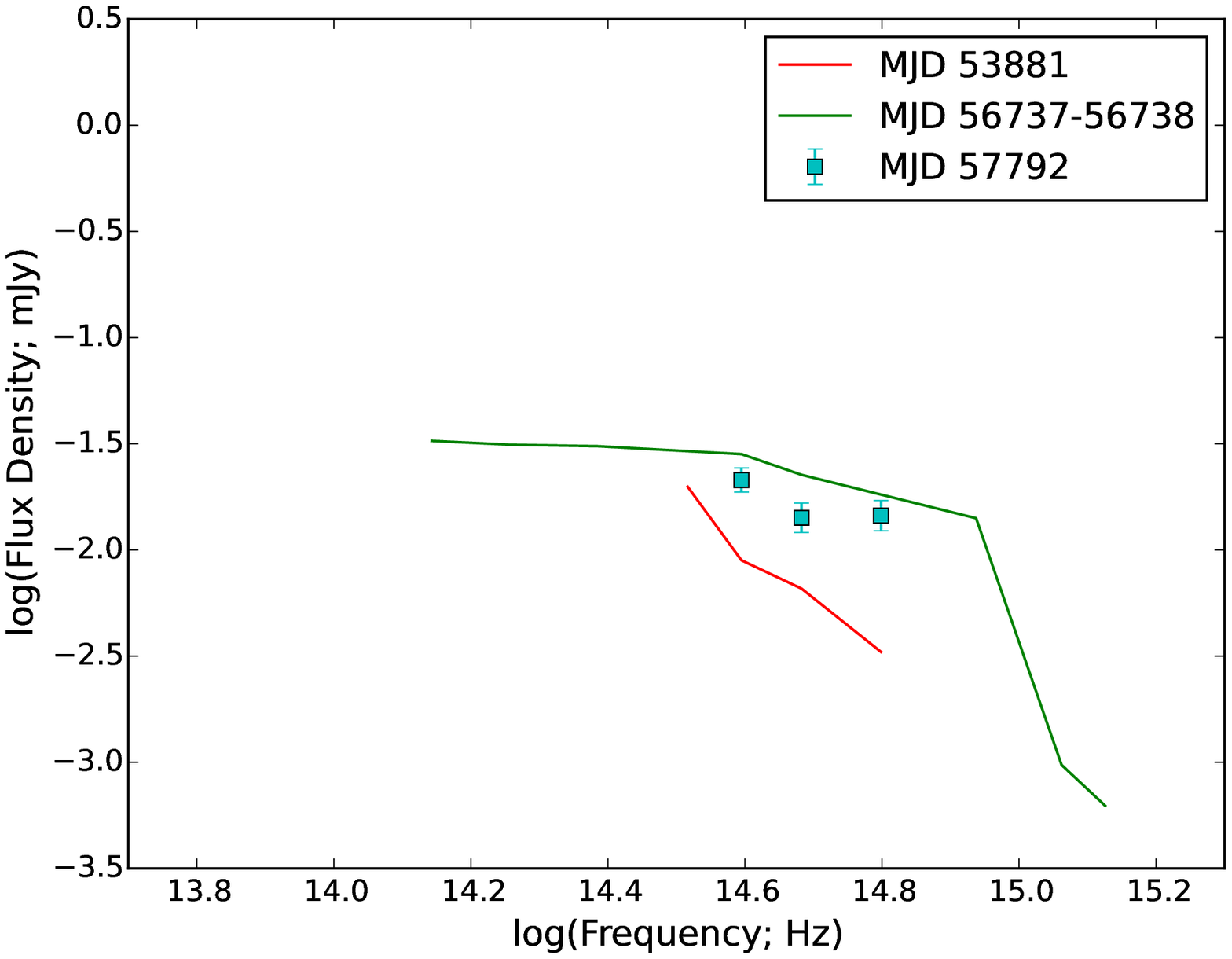}
\caption{Individual OIR SEDs from date ranges indicated in each panel. The SEDs appear in chronological order and show the evolution of the spectrum from 2006 to 2017. To aid the eye, the solid lines show representative SEDs when the spectrum was faint and red (red curve) and bright and flatter (green curve).}
\end{figure*}

On several dates (with data taken on the same MJD), magnitudes exist in two or more filters, so below we construct OIR SEDs in order to explore the evolution of the SED. On a number of dates only one or a few filters were used, but other observations with different filters were taken within some range of dates. In these cases, we construct SEDs over date ranges, spanning from 2 to 176 days. The evolution of the SEDs allows us to probe variability of the SED shape on timescales longer than the date ranges. The SED shape and how quickly it varies gives clues to the origin of the emission and high amplitude variability. Short-term variability \citep{shahet13} will cause some unavoidable scatter in the SEDs for any data that are not strictly simultaneous. Nevertheless, long-term evolution of the SED appears to be evident and is of higher amplitude (spanning one order of magnitude in flux) than the short-term variations \citep[which have a fractional rms of $\sim 35$\%;][]{shahet13}.

The data were de-reddened using the same method as in \cite{russet16}. The extinction value of $A_v = 0.124$ \citep{armaet13,corret16} was adopted, and the wavelength-dependent extinction terms are taken from \citet*{cardet89}. For the SEDs, we calculate the logarithm of the flux density, $log_{10}(F_{\nu}$; mJy$)$. The error on the log of the flux density is $\Delta (log_{10}(F_{\nu})) = 0.4 ~\Delta m$, where $\Delta m$ is the magnitude error.

\begin{figure}
\centering
\includegraphics[height=8.7cm,angle=270]{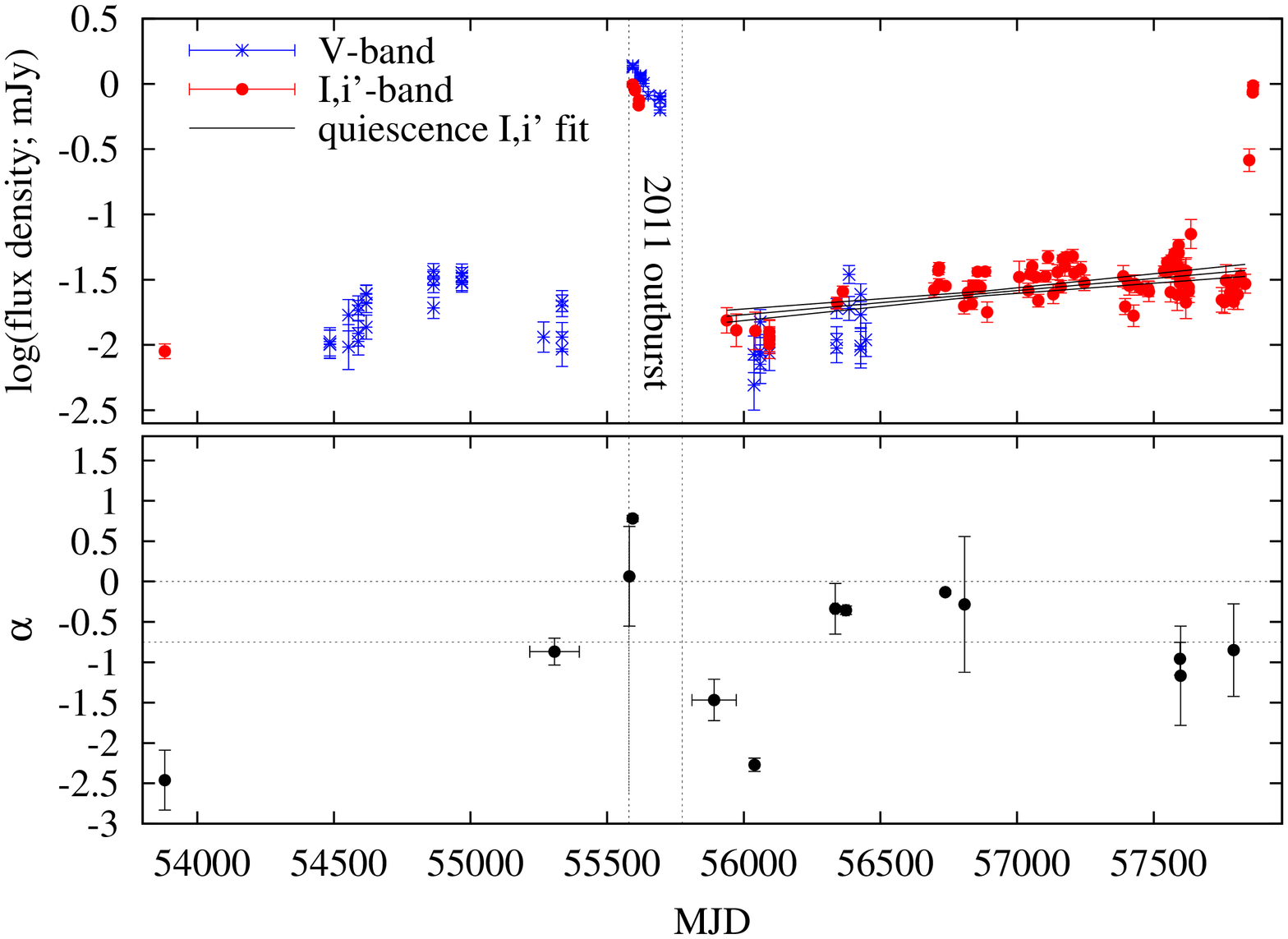}
\includegraphics[height=8.7cm,angle=270]{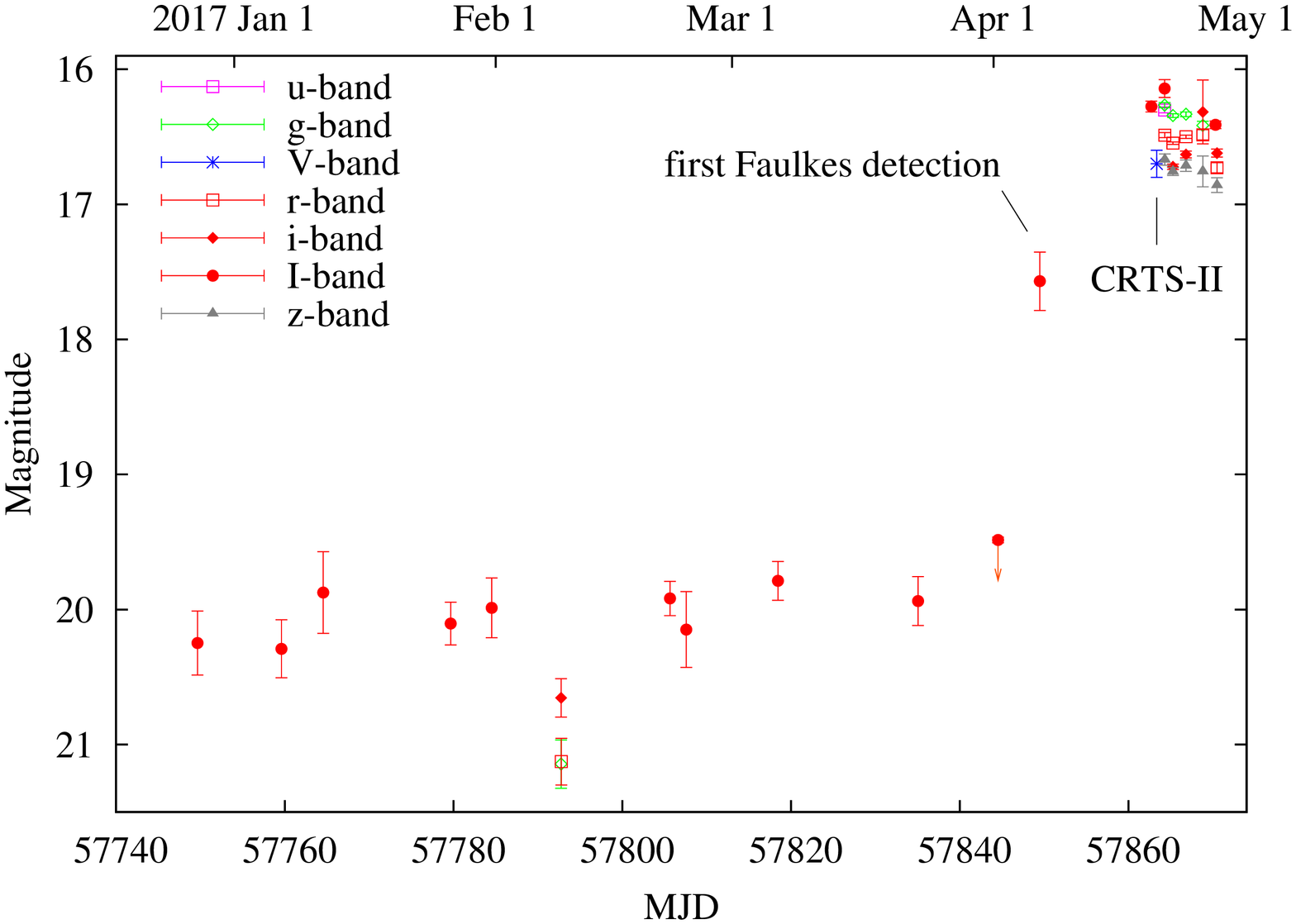}
\caption{Upper: long-term evolution of the de-reddened flux density (upper panel) and OIR spectral index (lower panel). The two horizontal dotted lines at $\alpha = 0$ and $\alpha = -0.75$ represent typical values for optically thick and optically thin synchrotron spectra, respectively. Lower: 2017 light curve (magnitudes), showing the rise of the 2017 outburst.}
\end{figure}

The resulting OIR SEDs are presented in Figs 2 and 3. The OIR spectrum of the source is clearly bluer ($\alpha > 0$) during outburst and redder ($\alpha < 0$) in quiescence, as first reported by \cite{shahet13}. During quiescence, the near-IR data (at $\log (\nu/$Hz) $< 14.4$) span a smaller range of flux densities than the optical and UV data ($\log (\nu/$Hz) $> 14.4$; Fig. 2). The near-IR data also appear flatter ($\alpha \sim 0$) than the optical ($\alpha < 0$) data. The evolution of the SED can be studied in Fig. 3, where each SED is shown in a separate panel, with two representative SEDs overlaid in solid lines for comparison. These two SEDs are from MJD 53881, the first data taken in 2006 in which the source was faint and red \citep{shahet13}, and from MJD 56737--56738, the well-sampled, quasi-simultaneous broadband SED presented in \cite{plotet16} when the source was brighter and consistent with a jet spectrum with a break at $\nu_{\rm b} \sim (2$--$5) \times 10^{14}$ Hz.

Two SEDs (MJD 55580--55581 and MJD 55593) show data from the rise and peak of the 2011 outburst. Before and soon after the outburst, the OIR SED appears red and steep. In particular, the SEDs between MJD 55810 and 56093 (2011 September--2012 June) are consistent with the SED acquired by SDSS in 2006 (MJD 53881). Then, between MJD 56093 and MJD 56362 (2012 June--2013 March) the SED appears to evolve, the optical fluxes brighten, and the spectrum becomes flatter. For the remainder of the SEDs (2013--2017), the OIR spectrum is close to the well-sampled jet break spectrum of \cite{plotet16} from MJD 56737--56738 (2014 March), with the spectrum becoming slightly redder in 2016--2017. The last quiescent SED was taken on 2017 February 8 (MJD 57792), just two months before the rise of the 2017 outburst.

The increase of optical flux in quiescence between 2011 and 2017 is quantified in Fig. 4 (upper panels). 
For the quiescent data after the 2011 outburst, we fit the log(flux density) $I,i^{\prime}$-band light curve (top panel) with a linear function and find a significant rise in flux over the 5.1 years of data. We measure a rise rate of $0.17 \pm 0.03$ mag yr$^{-1}$ (exponential rise in flux, linear rise in magnitude). Each SED in Fig. 3 is fitted with a single power law, and the resulting spectral index is shown in the next panel of Fig. 4. For the SED on MJD 56737--56738, we fitted the $i^{\prime}$-to-$K_{\rm s}$-band data only, because the fluxes dropped at higher frequencies than the jet break \citep[see][]{plotet16}. For all other dates, a single power law is capable of fitting the SED; a broken power law would generally result in large parameter errors due to the few number of data points. For the SEDs of MJD 56086--56093 and 56834--56841, the data errors are large, resulting in poor estimates of $\alpha$, and these were excluded from Fig. 4. We find that $\alpha$ varies between $\sim -2.5$ and 0 in quiescence, with a redder index when the source was faintest, such as soon after the 2011 outburst. Just prior to the 2017 outburst, the spectral index was $\alpha \sim -1$, consistent with optically thin synchrotron, and a similar spectral index was measured just prior to the 2011 outburst.

In the lower panel of Fig. 4 our 2017 monitoring is shown, with the rise into the 2017 outburst. The first clear detection of the outburst was on April 6, when the magnitude was $I = 17.6 \pm 0.2$; $>$2 mag brighter than all of the 2016 detections prior to that date. On April 1, the magnitude was $I > 19.49$ ($3\sigma$ upper limit), which is fainter than the brightest detection in quiescence, so the source was still in quiescence on April 1. The outburst therefore began between April 1 and 6, and must have brightened at a rate of $\geq 0.34$ mag/d in $I$-band.

\begin{table*}
\begin{center}
\caption{Optical rise rates of BH LMXBs in quiescence.}
\begin{tabular}{lccccl}
\hline
Source & \multicolumn{3}{c}{-------- Rise Rates (mag yr$^{-1}$) --------}  & Years of  & References \\
       & $V$ & $R$ & $I$ or $i^{\prime}$ & Data$^{a}$ & \\
\hline
H1705--250          &                   &                   & $0.083 \pm 0.022$ & 6.3  & \cite{yanget12} \\
V404 Cyg            & $0.048 \pm 0.009$ & $0.035 \pm 0.003$ & $0.022 \pm 0.002$ & 2.9  & \cite{bernet16} \\
GRS 1124--68        & $0.036 \pm 0.001$ &                   & $0.020 \pm 0.000$ & 11.2 & \cite{wuet16} \\
GS 1354--64         &                   & $0.088 \pm 0.006$ & $0.058 \pm 0.004$ & 6.8  & \cite{koljet16} \\
Swift J1357.2--0933 &                   &                   & $0.169 \pm 0.027$ & 5.2  & This paper \\
\hline
\end{tabular}
\normalsize
\end{center}
$^{a}$For V404 Cyg, GS 1354--64 and Swift J1357.2--0933 the optical rise precedes a new outburst of the source, whereas for H1705--250 and GRS 1124--68 no new outburst has yet been detected. The rise in V404 Cyg occurred after a long-term slow fade.
\end{table*}

\section{Discussion and Conclusions}

The optical/infrared SEDs are generally well described by a power law that evolves from a faint, red slope ($\alpha < -1$) before and soon after the 2011 outburst, to one which is bright and flat ($\alpha \sim 0$) or optically thin ($\alpha \sim -1$) since 2013. This transition occurs as the optical flux rises quite steeply. On dates when $\alpha < -0.5$, the jet break must have resided at frequencies lower than that sampled in the SED (i.e., in the infrared), whereas when $\alpha \sim 0$ the jet break shifted up to optical frequencies. We find that the evolving spectrum is responsible for the long-term brightening, and identify a transition between 2012 June and 2013 March during which the jet spectral break shifted from infrared to optical wavelengths and the optical flux brightened. The jet break then shifted back to the infrared by 2016--2017.

As a caveat, we cannot formally rule out a jet break shifting on shorter timescales than our sampling, for example hour timescales, as has been observed from GX 339--4 during outburst \citep{gandet11}. However, what we do observe is an OIR SED shape that is stable over months--years, which appears inconsistent with such dramatic variations in the spectrum on hour timescales. Additionally, we note that intrinsic source reddening changes cannot account for the high amplitude variability. On short timescales, the fractional rms variability is stronger at near-IR than optical wavelengths \citep[Figure 5 in][]{shahet13}, whereas the opposite dependency on wavelength would be expected if extinction were responsible. Although some dips were seen in the fast timing light curve of \cite{shahet13} that are analogous to the quasi-periodic dips seen during outburst by \cite{corret13}, which are caused by obscuration, most of the variability is not described by dips, and a different power density spectrum \citep[fig. 5 in][]{shahet13} would be expected. Likewise, if such dips dominated the long-term variability, one would expect most of the magnitudes during quiescence to be roughly the same, with a few fainter magnitudes that corresponded to dips. This is not a good description of the quiescent light curve. In addition, the interstellar reddening is very low; $A_v = 0.124$ \citep{armaet13,corret16}, and intrinsic dust obscuration would have to be local to the source which, given its high galactic latitude, also seems unlikely.

Few studies report long-term variations in the quiescent optical flux in LMXBs. In A0620--00, \cite{cantet08,cantet10} identified three optical states; passive, loop, and bright, and since its 1975 outburst, the source has made transitions between these states several times.
The change in magnitude was greater in the blue bands than in the red bands. 
Active and passive states were also reported from OIR monitoring of V4641 Sgr in quiescence \citep{macdet14}. In both sources, no long-term optical brightening or fading was found.

A long-term rise has been reported recently in four BH LMXBs. We compare the rise rate we have estimated for Swift J1357.2--0933 with these other systems in Table 1. The rise rates in $I$ or $i^{\prime}$-band are between 0.02 and $\sim 0.08$ mag yr$^{-1}$ for all sources except Swift J1357.2--0933, which has a much higher rise rate of $\sim 0.17$ mag yr$^{-1}$. For the three sources with measurements in more than one filter, the rise is greater in the bluer filters (shorter wavelengths), as was similarly found for the active state of A0620--00 \citep{cantet08}, and as may be expected as the accretion disk temperature increases, making the OIR SED bluer. We do not have well-sampled light curves of Swift J1357.2--0933 in other filters; however, over the five years between outbursts there is a $u$-band flux increase of a factor of $\sim 14$, an $I$-band flux increase of a factor of $\sim 4$--6, and almost no change at all in the infrared flux. The rise is therefore greater at shorter wavelengths. 
However, the SED evolution presented in Figs 2 and 3 indicate that, for Swift J1357.2--0933, the bluer-when-brighter behavior is caused by an evolving synchrotron jet spectrum (and not a heating accretion disk, or indeed, local reddening changes).

Swift J1357.2--0933 is a short orbital period (2.8 hr), high-inclination BH LMXB that exhibited optical dips during its 2011 outburst due to quasi-periodic obscuration of the accretion flow \citep{corret13}. Because the accretion disk is small and viewed almost edge-on, the disk emission reaching the observer must necessarily be reduced compared to lower inclination and longer period systems. The projected disk surface area along our line of sight toward the source is therefore much smaller than almost all other BH LMXBs. This could explain why in Swift J1357.2--0933 the jet (if relativistic beaming does not play a significant role), and not the disk, appears to dominate the OIR emission in quiescence \citep[although emission lines from the disc have been detected in quiescence in 2014 by][]{mataet15}. If this is the case, we may expect similar high amplitude variability, red SED and polarization from other high-inclination, short period systems. The companion star must also be dimmer than the jet and have a magnitude of $V \gtrsim 22$ to explain the high amplitude variability and SEDs. It is worth noting that all LMXBs that have radio detections in the quiescent state are close-by sources, whereas Swift J1357.2--0933 is likely to lie at a further distance, which makes the OIR jet detections more important for this system. In addition, the high galactic latitude of the source has allowed sensitive UV observations, which have helped to determine the unusually steep OIR spectral index. These UV detections in quiescence are impossible for most BH LMXBs due to their typically higher extinction.

The rise rate of Swift J1357.2--0933 is the closest to that predicted by the DIM \citep{dubuet01,laso01}, even though in the DIM the disk is producing the optical emission, not the jet. The jet luminosity is likely driven by the mass accretion rate in the inner regions of the accretion flow. The radio--X-ray correlation in BH X-ray binaries extends to quiescence \citep{gallet14,plotet17}, so the two emission processes are likely to be linked, and we speculate that the relatively fast rise seen in Swift J1357.2--0933 is due to the increase in the mass accretion rate onto the BH preceding the outburst (most noticeably during the transition in 2012--2013). Indeed, the DIM predicts an optical brightening between outbursts due to the increase of mass accretion rate at the inner edge of the disk, which, to have long recurrence times, must be truncated between outbursts. In low-inclination systems, the outer disk emitting in the optical may not be as good a tracer of this rise, because the outer disk is never depleted between outbursts, whereas the inner regions are. Therefore in most systems, the gradual increase in disk temperature predicted by the DIM could be occurring at a characteristic radius that is smaller than the bulk of the optical emission, producing only a modest optical rise between outbursts. 
For such low-inclination sources, Table 1 suggests that a long-term optical rise does appear to precede outbursts, and so outbursts of H1705--250 and GRS 1124--68 could be expected in the next few years. This emphasizes the importance of OIR monitoring of quiescent LMXBs, in particular to identify long-term flux increases that could be the precursor to a forthcoming outburst. The edge-on source Swift J1357.2--0933 is the best example known because of its clear steep rise, and considering it is also likely the only quiescent source in which the optical jet properties can be regularly monitored, until fainter sources are visible regularly, which will be the case when the Large Synoptic Survey Telescope \citep[LSST;][]{lsst09} is in operation.

\acknowledgments

We thank Poshak Gandhi for excellent suggestions that have improved the discussion. R.M.P. acknowledges support from Curtin University through the Peter Curran Memorial Fellowship. This project has received funding from the European Union's Horizon 2020 research and innovation programme under the Marie Sklodowska-Curie grant agreement no. 664931. The Faulkes Telescopes are maintained and operated by the Las Cumbres Observatory (LCO).

\vspace{5mm}
\facilities{LCO:Faulkes 2m, 1m.}
\software{IRAF \citep{tody86,tody93},
GNUPLOT \citep{jane16}
          }

\end{document}